\documentclass[twocolumn,showpacs,preprintnumbers,amsmath,amssymb]{revtex4}

\usepackage{graphicx}
\usepackage{dcolumn}
\usepackage{bm}

\begin{document}
\title{Heavy Majorana neutrinos at a very large electron-proton collider.}
\author{F.M.L. de Almeida Jr.}
\email{marroqui@if.ufrj.br}
\author{Y. A. Coutinho}
\author{J. A. Martins Sim\~oes}
\author{M.A.B. do Vale}
\altaffiliation[Now at ] {Natural Sciences Department, FUNREI, MG, 
Brazil}
\affiliation{Instituto de F\'\i sica\\
Universidade Federal do Rio de Janeiro, RJ, Brazil \\}
\date{\today}
\begin{abstract}
\par
We study the discovery potential for detecting new neutral heavy Majorana leptons as suggested by some extentions of the Standard Model in recently proposed electron-proton colliders. Since 1998-1999 the option of an electron-proton collider for the Very Large Hadron Collider at Fermilab operating with the  proton booster has been considered. We consider the reaction $ e^{-}p \longrightarrow e^{+}+ jets$ and present estimates for the signal and Standard Model background including hadronization. 
\end{abstract}
 
\maketitle
\section{Introduction}
The recent experimental results on solar and atmospheric neutrino properties \cite{NIS} confirm that standard neutrinos have small masses and oscillate. However, the nature of those massive neutrinos - if they are Dirac or Majorana particles - is still an open question. This is connected with lepton number conservation and to the general symmetries of some  extention of the standard model. The smallness of neutrino masses is generally understood as a consequence of the see-saw mechanism. This brings the question of the possibility of new heavy neutrino states. So far none of these new states was experimentally observed \cite{PDG,ZUB} with masses up to $M_N\simeq$ 100 GeV. For higher masses there are many suggestions of experimental possibilities in the next high energy hadron-hadron colliders \cite{YSP,ALI}, in electron-positron linear accelerators \cite{ARS,DJO,ACV}, in neutrinoless double-beta decay \cite{BIL}. Another experimental possibility is found on electron-proton colliders with a positron in the final state \cite{BUM}. This means a very clear experimental signature that violates the electron leptonic number by two units. In this paper we investigate the discovery potential for recently proposed electron- proton colliders for detecting new neutral heavy Majorana leptons as suggested by extentions of the Standard Model. Since 1998-1999 the option of an electron-proton collider for the Very Large Hadron Collider at Fermilab operating with the proton booster has been considered. We study the reaction $ e^{-}p \longrightarrow e^{+}+$ jets and present estimates for the signal and Standard Model background.

\section{The Very High Energy Electron-Proton Collider}

There is a growing work on the options for a future Very Large Hadron Collider
(VLHC)\cite{SYP}. The low magnetic field (approximately 2 T) collider design in a tunnel can also be used for a future high magnetic field (approximately 10 T) collider. These designs could also be used for electron-proton colliders\cite{VLH,DER}. In the low-field version electron-proton 
collisions would be produced by an $80$ GeV electron ring and a $3$ TeV proton 
booster ($\sqrt{s} = 980$ GeV), with a luminosity ${\cal L}$= 2600 pb$^{-1}$/yr, 
while in the high-field option one could reach a center-of-mass energy $\sqrt{s} = 
6320$ GeV and a luminosity  ${\cal L}$  = 1400 pb$^{-1}$/yr with the collision of a $200$ GeV electron
beam with a $50$ TeV proton beam. These two options are considered in the present work.
\par
\section{The model}
New heavy leptons are present in many possible extentions of the standard model. In the superstring inspired scenario \cite{SIR} we have new doublet vector leptons. In mirror models \cite{MIR} there are new fermions with right handed doublets and left handed singlets. Some see-saw extentions of the standard model include new weak-singlet fermions \cite{BET}. 
\par
The general mixing  of light and heavy Majorana neutrinos has new mixing parameters as well as new possible CP violating phases\cite{FRI}. It is well known that in neutrinoless double beta decay the theoretical estimates depend strongly on mixing angles and CP phases \cite{BEL}. As a consequence, the comparison between the experimental limits and theory is model dependent. This is also the case for other processes. In this paper we are interested in the study of the discovery potential of new electron-proton colliders for new heavy Majorana neutrinos. From the experimental point of view,instead of start comparisons of very particular models with a great number of parameters and hypothesis one must start looking for  the lightest of the new possible Majorana neutrinos, coupled to the electron family. This   reduces the number of unknown parameters to two; the heavy neutrino mass and one new mixing angle, that should be determined experimentally.
\par
In this paper we will adopt a general parametrisation that applies to all the above models.
We will suppose that mixing angles for heavy-to-light neutrinos and new heavy neutrino masses are independent parameters \cite{BUC}. The light neutrinos couplings to the neutral Z are given by $ g_{V,A}=g_{V,A}^{SM} - {sin^2{\theta_{\mbox {mix}}}/2}$. Taking the experimental values for $g_{V,A}$ and the Standard Model predictions we obtain a small upper bound for $\theta_{\mbox {mix}}$. A recent estimate \cite{ACV} gives $\sin^2 \theta_{\mbox{mix}} < 0.0052$ with $95 \%$ C.L. This limit value is used throughout this paper for all curves and distributions. The decay of the Majorana neutrino into muons and taus are also possible. In ref. \cite{ACV} it was found that for the muon sector the experimental mixing angle limit with $95\%$ C.L. is much smaller than for the electron sector.
\par
Most extended models predict new fermions and new gauge vector bosons. Since we presently have no signal for new interactions, we will make the hypothesis that the new heavy neutrino states behave as $SU_L(2)\otimes U_Y(1)$ basic representations and interact mostly with the standard model gauge bosons. We can resume the new particle  interactions in the neutral and charged current lagrangians:

\begin{equation}
{\cal L}_{nc}=-\frac{g}{4c_W } {sin{\theta_{mix}}} Z_{\mu}\overline{\psi_N}\gamma^{\mu}
\left(1-\gamma_{5}\right)\psi_{\nu} + h.c..
\end{equation}

and

\begin{equation}
{\cal L}_{cc}=-\frac{g}{2\sqrt{2}} {sin{\theta_{mix}}}
W_{\mu}\overline{\psi_N}\gamma^{\mu}\left(1-\gamma_5\right)\psi_e + h.c.
\end{equation}
and similar terms for the other leptonic families. In the Majorana neutrino case, each completely neutral heavy lepton is coupled to all charged leptons but we are considering only the lower mass state.

\par
The decay modes for these leptons, in the Majorana case \cite{YSP} must include both signatures $N \longrightarrow \ell^{\mp}\, W^{\pm}$  and $ N \longrightarrow \nu_{\ell}\,(\bar\nu_{\ell})\, Z$, with $\ell=e,\mu,\tau$. In this work we have considered the Majorana neutrino decay only to the first family since for the other families we must add more mixing parameters. The Majorana neutrino branching ratios are shown in Table I and explicit formulas for the widths are given in ref. \cite{PAN}

\begin{table}
\caption{
\label{tab:table1} 
Branching ratios for $N$ decays in the first family.}
\begin{center}
\begin{tabular}{|c|c|c|}\hline
Decays & $M_{N}=100$ GeV & $M_{N}=1500$ GeV \\ \hline
$N \rightarrow e^+ W^-$ & 44$\%$ & 33$\%$ \\ \hline
$N \rightarrow e^- W^+$ & 44$\%$ &  33$\%$ \\ \hline
$N \rightarrow \nu_e Z$ & 5,8$\%$ & 17$\%$ \\ \hline
$N \rightarrow \bar\nu_e Z$ &  5,8$\%$  & 17$\%$ \\ \hline
\end{tabular}
\end{center}
\end{table}

\section{Results}

\begin{table*}
\caption{\label{tab:table2}Number of events and 
significance for heavy Majorana neutrino production.}
\begin{center}
\begin{tabular}{|c|c|c|c|c|c|}\hline
 $\sqrt s$ (GeV) & $M_{N}$ (GeV) & $\Gamma_N (GeV)$ &Events after basic cuts &
 Events after all cuts($\%$) & $S/\sqrt{S+B}$ \\ \hline
 980 & 100   & 1.13 $10^{-3}$ & 516   &  65 (12$\%$)  & 8  \\ \cline{2-6}
     & 300   & 1.38 $10^{-1}$ & 83   &  32 (38$\%$)  & 5.6  \\ \hline
 6320 & 500  & 6.50 $10^{-1}$ & 334   &  153 (45$\%$)  & 12.4 \\ \cline{2-6}
      & 1500 & 17.6 $10^{0}$ & 80    &  24 (30$\%$)  &  5  \\ \hline
\end{tabular}
\end{center}
\end{table*}

In Fig. 1 we show the structure of the Feynman diagrams that contribute for the signal. We have included in our results all the Majorana off-shell and on-shell contributions. The calculations for the signal were done using the high energy package CompHep \cite{HEP} with CTEQ95 structure functions, after implementation of the extended model. We call attention to the fact that the N width is very narrow for lower heavy neutrino masses and increase for higher masses ( see Table II). Thus propagator effects must  be handled with care. 

\par
The cross sections for on and off-shell heavy Majorana neutrino production as a function of its mass are displayed in Fig. 2 for $ \sqrt{s}=$ 980 GeV and Fig. 3 for $ \sqrt{s}=$ 6320 GeV. In all our results we have done first general detector cuts for the final positron $E_{e^+}> 5$ GeV and $\vert \eta_{e^+} \vert < 3.5$, where $\eta_{e^+}$ is the pseudo rapidity defined relative to the initial proton. 
\par
In order to reproduce the signal and background as close as possible to experimental data we have employed the package PYTHIA 6.1 \cite{PIT} to generate the quark hadronization and decay from the Majorana neutrino production as well as the corresponding standard model background. The signal events were generated first at parton level with the CompHep package as above mentioned and treated as external process to the PYTHIA program. In this way we have simulated the W, Z, quark and hadron  decays. The Standard Model background was estimated leaving open all the possible channels in the PYTHIA program. The positrons come mainly from charm decay. In both cases we did not allow pions to decay.
\par
In the final visible particles we have imposed the following criteria to select events:
\par                      

a) For $\vert \eta_{e^+} \vert < 3.5$ and $E_{e^+}> 5$ GeV we must have the number of $e^{+} > 0 $,
\par
b)For $\vert \eta_{\, h_i} \vert < 3.5$ and $E_{h_i}> 5$ GeV we must have the number of hadrons $> 5$.
 \par
With these two conditions, more than 99 $\%$ of the background events were eliminated for both energies.
\par
In order to reconstruct the Majorana neutrino invariant mass we have employed the algorithm cluster PYCELL from the PYTHIA package, with 60 azimuthal divisions and 26 divisions in the electron beam direction. For $\sqrt s$= 980 GeV  we have applied the condition $ 0.7 < \sqrt{(\Delta \eta)^2 + (\Delta \phi)^2}$ while for $\sqrt s$= 6320 GeV  this condition was changed to $0.3 < \sqrt{(\Delta \eta)^2 + (\Delta \phi)^2} $ with $ \vert \eta \vert < 3.5$ for both energies. 
 \par
After these conditions, we have further demanded that:
\par
c) Number of clusters $>$ 1,
\par
d) Number of clusters with 1 track = 1,
\par
e) Cluster invariant mass in the range (70, 90) GeV.
\par
This last requirement was done to identify the W coming from the heavy Majorana neutrino decay. After these additional cuts, the background estimated for 650 k events for $ \sqrt s \simeq 1$ TeV and 2,4 M events for $ \sqrt s \simeq 6$ TeV was reduced to practically zero. This point was explicitly verified by generating 10 times more events than expected by the luminosity. For the signal, the same cuts imply the results shown in Table II. We also show in the last column of Table II. the significance $S/\sqrt{S+B}$ for each case, with "S" the number of signal events and "B" the number of background events.
\par
If we choose the clusters obtained with the surviving events then we can reconstruct the Majorana neutrino invariant mass as shown in Figs. 4 and 5.
\section{Conclusions}
The present work shows that heavy Majorana neutrinos could be produced in the electron-proton option for the VLHC. The very clear signature of a positron in the final state allows to separate efficiently the signal events from the background and to reconstruct the Majorana neutrino invariant mass. In the $ \sqrt s \simeq 1$ TeV option we can reach Majorana masses up to 300 GeV. For the $ \sqrt s \simeq 6$ TeV option this value could be as high as 1500 GeV. This is a conservative estimate since we are taking as limiting value $S/\sqrt{S+B} >5$.

\begin{acknowledgments} This work was partially supported by the following Brazilian agencies: CNPq and FAPERJ.
\end{acknowledgments}

\bigskip 
\LARGE
Figure Captions
\normalsize
\begin{enumerate}
\item General Feynman graphs for single heavy Majorana  contribution to  $ e^{-}p \longrightarrow e^{+} + W + q$.
\item Cross section for single heavy Majorana neutrino production at $\sqrt s=$ 980 GeV using basic cuts  $E_{e^+}> 5$ GeV and $\vert \eta_{e^+} \vert < 3.5$.
\item Same as figure 2 for  $\sqrt s=$ 6320 GeV.
\item Reconstruction of the invariant Majorana mass distribution for $\sqrt s=$ 980 GeV after hadronization  using cuts and PYCELL algorithm.
\item Same as figure 4 for $\sqrt s=$ 6320 GeV.
 \end{enumerate}
\end{document}